\documentclass[pra,twocolumn,showpacs,showkeys]{revtex4}

\usepackage{amssymb, amsmath, amsfonts, latexsym, verbatim, mathrsfs}

\usepackage{graphicx}

\newtheorem{theorem}{Theorem}

\begin{document}

\title{Relaxation to equilibrium driven via indirect control in Markovian dynamics}

\thanks{Work supported by the European grant ERG:044941-STOCH-EQ. Partially
supported by R. Zecchina (ICTP), to whom the author is grateful, and
by INFN, Sezione di Trieste, Italy. The author acknowledges F.
Benatti and R. Floreanini for many helpful discussions.}

\author{Raffaele Romano}
\email{rromano@ts.infn.it}

\affiliation{The Abdus Salam
International Centre for Theoretical Physics \\
Strada Costiera, 11 I-34014 Trieste, Italy}

\affiliation{The Department of Theoretical Physics, University of
Trieste, \\
Strada Costiera, 11 I-34014 Trieste, Italy}


\begin{abstract}

\noindent We characterize to what extent it is possible to modify
the stationary states of a quantum dynamical semigroup, that
describes the irreversible evolution of a two-level system, by means
of an auxiliary two-level system. We consider systems that can be
initially entangled or uncorrelated. We find that the indirect
control of the stationary states is possible, even if there are not
initial correlations, under suitable conditions on the dynamical
parameters characterizing the evolution of the joint system.

\end{abstract}



\maketitle


\section*{INTRODUCTION}

In the past decades quantum mechanical systems have attracted lot of
attention for their peculiar properties, that indicate they are good
candidates for the implementation of outperforming technologies in
the fields of information and computation\cite{niel}. In this
spirit, some amazing protocols have been recently developed, as for
example the computational algorithm for the factorization of a large
number \cite{shor}, or the schemes for teleportation \cite{benn1}
and quantum cryptography \cite{benn2}. Many concrete physical
systems have been proposed for the practical implementation of these
ideas, as optical devices, cold trapped atoms, nuclear spins in
magnetic fields (NMR), or quantum dots in electromagnetic cavities.

In all these cases, the largest obstacle to the implementation of
stable and efficient schemes is represented by the unavoidable
interaction of microscopic systems with the surrounding environment.
Because of this interaction, the system dynamics is subject to loss
of coherence, irreversibility and dissipation, and the appealing
properties of quantum systems are usually lost or compromised during
the time evolution.

Usually, the environmental action is accounted for by describing the
dynamics of the system $S$ through a Markovian one-parameter family
of maps $\{ \gamma_t; t \geqslant 0\}$, satisfying the semigroup
property $\gamma_{t + s} = \gamma_t \circ \gamma_s$, with $t, s
\geqslant 0$, with
\begin{equation}\label{eq00}
    \rho_S (t) = \gamma_t [\rho_S (0)],
\end{equation}
where the statistical operator (or density matrix) $\rho_S$ is an
Hermitian, positive, unit trace operator, acting on the Hilbert
space associated to the system, and representing its state. This
representation of the dynamics is not the most general, but it is
well justified in many cases, in particular when the coupling
between $S$ and the environment can be considered weak. The
generator $L$ of the dynamics can be obtained by writing
(\ref{eq00}) in differential form, $\dot{\rho}_s = L[\rho_S]$, and
it has the standard structure
\begin{equation}\label{eq01}
    L[\rho_S] = -i [H_S, \rho_S] + \sum_{i,j} c_{ij} \Bigl(
    F_i \rho_S F_j^{\dagger} - \frac{1}{2} \{ F_j^{\dagger} F_i, \rho_S \}
    \Bigr),
\end{equation}
where $H_S = H_S^{\dagger}$ is the system Hamiltonian, and the set
$\{F_i; i \}$ satisfies ${\rm Tr} F_i = 0$, ${\rm Tr} (F_i
F_j^{\dagger}) = \delta_{ij}$. The Kossakowski matrix $C = [c_{ij}]$
must satisfy $C^{\dagger} = C \geqslant 0$ in order to guarantee the
complete positivity of the evolution, and then the physical
consistency of the formalism \cite{alic,breu}. It encodes the
microscopical details of the interaction between system and
environment. The first term in the right hand side of (\ref{eq01})
represents the coherent part of the evolution, and the generator of
the system dynamics has this form whenever the interaction with the
surrounding environment can be neglected. The corresponding time
evolution is given by a group of reversible, unitary
transformations. The second term is responsible for irreversibility
and dissipation, since it produces a contraction map on the set of
states, and, in some cases, relaxation to stationary states.

It is of fundamental relevance to study methods to fight against
decoherence. When the environmental noise exhibits some particular
symmetry properties, this task can be realized by encoding the
relevant information in suitable {\it Decoherence Free Subspaces or
Subsystems} unaffected by decoherence (for a review, see
\cite{lida}). An active approach consist in directly affecting the
system dynamics, in order to preserve its relevant properties or
induce arbitrary manipulations. This {\it controlled evolution} is
realized through some functions, entering the dynamics, that can be
manipulated via external actions (see for example
\cite{tarn,rama,viol1,schi,viol2,albe,alta} for a geometric approach
to controllability).

Several approaches to the control of a quantum system have been
proposed in the past years. In the {\it open loop} schemes the
control functions are a priori fixed (that is, they are independent
on the state of the system). Conversely, in the {\it closed loop}
control schemes, the control functions are updated in real time by
feeding back some information about the actual state of the system,
usually gained via an indirect continuous measurement ({\it quantum
feedback} \cite{bela,wise,manc}).

The control functions usually affect the Hamiltonian of the system
$H_S$, since the environmental action is usually uncontrollable.
This approach is called {\it coherent control}, as it affects the
coherent part of the dynamics. Motivated by different experimental
scenarios, another control scheme has been introduced, in which an
auxiliary system is used to manipulate the target system through
their mutual interaction. This {\it indirect control} scheme is of
relevance whenever the system dynamics cannot be directly accessed
\cite{mand,roma1,fu}. It represents a complementary approach to
controllability, with interesting features concerning the
purification of mixed states \cite{roma3}, and when applied to the
dynamics of open systems \cite{roma2}, since it makes use of the
correlations between the two subsystems, that can be created by the
environmental action (described for the first time in \cite{brau}).

One of the unwished consequences of the environmental action on the
system dynamics is, in many cases, the collapse of the system into a
-in many cases, unique- equilibrium state, with a consequent
reduction of the reachable sets, and loss of control. In this work
we address the following question: is it possible to modify the
stationary states of a {\it target} system $T$, evolving under a
quantum dynamical semigroup, by means of an open-loop indirect
control? In other words, we introduce an auxiliary system, a quantum
{\it probe} $P$, couple it to $T$ and consider the evolution of the
joint system $S = T + P$, and finally discard $P$ by taking into
account only the degrees of freedom of $T$. Assuming that $S$ is
still described by a quantum dynamical semigroup, we study the
stationary states of the system $T$ alone, affected by $P$ through
the correlations between the two systems. The impact of both initial
correlations, and correlations created during the joint evolution,
is taken into account.

The plan of this work is the following. In Section \ref{sec0} we
review some algebraic tool for the determination of the stationary
states of a quantum dynamical semigroup. In Section \ref{sec1} we
specify the dynamical settings considered in this paper and we
derive the relevant algebraic quantities, introduced in Section
\ref{sec0}. In Section \ref{sec2} we describe all the possible
scenarios for the stationary states, in terms of the dynamical
parameters characterizing the semigroup. In Section \ref{sec3} we
summarize our results, describe their physical significance and
finally conclude.


\section{Stationary states of quantum dynamical semigroups}
\label{sec0}

In general, the second contribution in the right hand side of
(\ref{eq01}) leads to the appearance of attractors in the state
space of $S$, and consequently relaxation to equilibrium of the
states of the system, absent if there is not interaction with the
environment. A stationary state for the dynamics, $\rho^{\infty}_s$,
is determined by the condition on the generator $L[\rho_S^{\infty}]
= 0$. This is a system of linear equations that can be solved using
standard algebraic tools. In this spirit, quantum dynamical
semigroups have been classified in terms of their relaxing
properties \cite{lend}. In the {\it uniquely relaxing semigroups},
there is a unique stationary state, and every initial state
eventually collapses to it. In the {\it relaxing semigroups},
although every trajectory collapses to a fixed state, this state is
not unique but depends on the initial conditions. Finally, in the
{\it non-relaxing semigroups}, oscillatory solutions survive. Even
if this method is very general, the resulting algebraic equations
are complicate, therefore we will rely on a different approach.

For the Markovian dynamics (\ref{eq01}), necessary conditions for
the existence of stationary states and for the convergence of
$\rho_S (t)$ to them have been derived in terms of the operators $\{
V_i; i \}$ appearing in the diagonal form of (\ref{eq01}),
\begin{equation}\label{eq02}
    L[\rho_S] = -i [H_S, \rho_S] + \sum_{i} \Bigl(
    V_i \rho_S V_i^{\dagger} - \frac{1}{2} \{ V_i^{\dagger} V_i, \rho_S \}
    \Bigr).
\end{equation}
The following theorem summarizes these conditions \cite{frig}, and
it will be the basis of our analysis.

\begin{theorem}
\label{theo1} Given the quantum dynamical semigroup (\ref{eq02}),
assume that it admits a stationary state $\rho_0$ of maximal rank.
Defining $\mathcal{M} = \{ H_S, V_i, V_i^{\dagger}; i\}^{\prime}$,
the commutant of the Hamiltonian plus the dissipative generators,
and $I$ the identity operator, the following conditions hold true:

    1. If $\mathcal{M} = span (I)$, then $\rho_0$ is the
    unique stationary state. Moreover, if $\{ V_i; i \}$ is a self-adjoint
    set with $\{ V_i; i \}^{'} = span (I)$, then for every
    initial condition $\rho_S (0)$
    $$\lim_{t \rightarrow + \infty} \rho_S (t) = \rho_0.$$

    2. If $\mathcal{M} \ne span (I)$, then there exist a
    complete family $\{ P_n; n \}$ of pairwise orthogonal projectors
    such that $\mathcal{Z} = \mathcal{M} \cap \mathcal{M}^{'} = \{ P_n; n
    \}^{''}$. If $\{ V_i; i \}^{'} = \mathcal{M}$, two extreme cases together
    with their linear superpositions may occur.
    If $\mathcal{Z} = \mathcal{M}$, then for every initial condition $\rho_S (0)$
    $$\lim_{t \rightarrow + \infty} \rho_S (t) = \sum_n Tr \bigl(P_n \rho_S (0) P_n\bigr)
    \frac{P_n \rho_0 P_n}{Tr (P_n \rho_0 P_n)}.$$
    If $\mathcal{Z} = \mathcal{M}^{'}$, then for every $\rho_S (0)$
    $$\lim_{t \rightarrow + \infty} \rho_S (t) = \sum_n P_n \rho_S (0) P_n.$$
\end{theorem}

Therefore, in order to characterize the stationary states of a
quantum dynamical semigroup, it is necessary to find a maximal rank
stationary state $\rho_0$, and to evaluate the algebras ${\mathcal
M}$ and ${\mathcal M}^{\prime}$.


\section{Dynamical settings and relevant algebras}
\label{sec1}

We assume that $S = T + P$ is a bipartite system, where $T$ and $P$
are two copies of the same two-level system, separately interacting
with a common environment according to the Markovian dynamics
(\ref{eq01}). The operators $F_i$ are given by $F_i = \sigma_i
\otimes \mathbb{I}$ for $i = 1, 2, 3$ and $F_i = \mathbb{I} \otimes
\sigma_{i-3}$ for $i = 4, 5, 6$, where $\mathbb{I}$ is the
2-dimensional identity operator, and $\sigma_i$, $i = 1, 2, 3$ are
the Pauli operators. We consider the standard representation of
these operators in which $\sigma_3$ diagonal. The matrix $C$ has the
form
\begin{equation}\label{eq03}
    C = \left[%
\begin{array}{cc}
  A & B \\
  B^{\dagger} & A \\
\end{array}%
\right],
\end{equation}
where $A = A^{\dagger}$ is the Kossakowski matrix for the system $T$
(or $P$) alone, and $B$ represents the dissipative coupling between
the two parties. The form (\ref{eq03}) is not the most general joint
Kossakowski matrix. More complicate expressions should be taken into
account when three-body contributions are relevant (common
interactions between the two subsystems and the surrounding), and
when the dissipative couplings of $T$ and $P$ to the environment are
different (for example, in a non-homogenous medium). We will limit
our attention to models satisfying (\ref{eq03}); moreover, for
simplicity, we will further assume $B = B^{\dagger}$. This
assumption highly simplifies the mathematical formalism.

Following Theorem \ref{theo1}, we need to write $C$ in diagonal form
in order to find the operators $V_i$ appearing in (\ref{eq02}). This
is achieved by means of the unitary transformation
\begin{equation}\label{eq04}
    U C U^{\dagger} = {\rm diag} (\lambda_i, i = 1, \ldots, 6),
\end{equation}
where $\lambda_i$ are the eigenvalues of $C$, $U$ is of the form
\begin{equation}\label{eq05}
    U = \frac{1}{\sqrt{2}} \left[%
\begin{array}{cc}
  \tilde{U} & \tilde{U} \\
  - \hat{U} & \hat{U} \\
\end{array}%
\right]
\end{equation}
and $\tilde{U}$, $\hat{U}$ are unitary transformations such that
\begin{eqnarray}\label{eq06}
    \tilde{U} (A + B) \tilde{U}^{\dagger} &=& {\rm diag} (\lambda_i^+, i = 1, 2,
    3), \nonumber \\
    \hat{U} (A - B) \hat{U}^{\dagger} &=& {\rm diag} (\lambda_i^-, i = 1, 2,
    3).
\end{eqnarray}
The eigenvalues of $C$ are ordered as $\lambda_i = \lambda^+_i$ for
$i = 1, 2, 3$ and $\lambda_i = \lambda^-_{i-3}$ for $i = 4, 5, 6$.
Comparing the generator forms (\ref{eq01}) and (\ref{eq02}), and
using the notation $U = [u_{ij}]$, we have
\begin{equation}\label{eq07bis}
V_i = \sqrt{\lambda_i} \sum_{k = 1}^6 u_{ik}^* F_k, \qquad i = 1,
\ldots,6.
\end{equation}
Following (\ref{eq05}), it is possible to write
\begin{equation}\label{eq07}
    \frac{1}{\sqrt{\lambda_i}} V_i = \left\{%
\begin{array}{ll}
    \mathbb{I} \otimes \tilde{\sigma}_i + \tilde{\sigma}_i \otimes \mathbb{I}, & \hbox{{\it i} = 1, 2, 3}
    \\ \\
    \mathbb{I} \otimes \hat{\sigma}_{i - 3} - \hat{\sigma}_{i - 3} \otimes \mathbb{I}, \quad & \hbox{{\it i} = 4, 5, 6} \\
\end{array}%
\right.
\end{equation}
where we have defined
\begin{equation}\label{eq08}
    \tilde{\sigma}_i = \sum_{k = 1}^3 \tilde{u}^*_{ik} \sigma_k, \quad \quad
    \hat{\sigma}_i = \sum_{k = 1}^3 \hat{u}^*_{ik} \sigma_k,
\end{equation}
and we used the notation $\tilde{U} = [\tilde{u}_{ij}]$, $\hat{U} =
[\hat{u}_{ij}]$. The operators in (\ref{eq08}) satisfy ${\rm Tr} \,
\tilde{\sigma}_i = {\rm Tr} \, \hat{\sigma}_i = 0$ and ${\rm Tr}
(\tilde{\sigma}_i \tilde{\sigma}_j^{\dagger}) = {\rm Tr}
(\hat{\sigma}_i \hat{\sigma}_j^{\dagger}) = \delta_{ij}$. They are
self-adjoint if and only if the unitary operators $\tilde{U}$ and
$\hat{U}$ are orthogonal.

\noindent The commutant of Theorem \ref{theo1} can be expressed as
\begin{equation}\label{eq09}
    \{ H_S, V_i, V_i^{\dagger}; i \vert \lambda_i \ne 0 \}^{'} =
    \bigcap_{i \vert \lambda_i \ne 0} \{ V_i, V_i^{\dagger} \}^{'}
    \cap \{ H_S \}^{'},
\end{equation}
where only non-vanishing eigenvalues $\lambda_i$ have to be
considered, otherwise the corresponding $V_i$ do not appear in the
generator (\ref{eq02}). Moreover, for a given $i$,
\begin{equation}\label{eq10}
    \{ V_i, V_i^{\dagger} \}^{'} = \{ v \vert v \in \{ V_i \}^{'}, v^{\dagger} \in \{ V_i \}^{'}
    \},
\end{equation}
therefore we can limit our attention to the sets $\{ V_i \}^{'}$. We
find convenient to consider separately the two kinds of
contributions defined in (\ref{eq07}). To begin with, we consider a
fixed index $i$ such that $\lambda_i^+ \ne 0$, and assume that the
corresponding $\tilde{\sigma}_i$ is non-singular. In this case it
can be written as
\begin{equation}\label{eq12}
    \tilde{\sigma}_i = \tilde{\mu}_i R_i
    \sigma_3 R_i^{-1}
\end{equation}
where
\begin{equation}\label{eq13}
    R_i = R_i^{-1} = \frac{1}{\tilde{\nu}_i} \left[%
\begin{array}{cc}
  \tilde{u}_{i3}^* + \tilde{\mu}_i & \tilde{u}_{i1}^* - i \, \tilde{u}_{i2}^*
  \\ \\
  \tilde{u}_{i1}^* + i \, \tilde{u}_{i2}^* & - \tilde{u}_{i3}^* - \tilde{\mu}_i \\
\end{array}%
\right],
\end{equation}
and
\begin{equation}\label{eq13bis}
    \tilde{\mu}_i^2 = \sum_j (\tilde{u}^*_{ij})^2, \quad \tilde{\nu}_i =
    \sqrt{2 \tilde{\mu}_i \left( \tilde{u}_{i3}^* + \tilde{\mu}_i
    \right)}.
\end{equation}
Since $\mathbb{I} \otimes \tilde{\sigma}_i + \tilde{\sigma}_i
\otimes \mathbb{I} = \tilde{\mu}_i \mathcal{R}_i ( \mathbb{I}
\otimes \sigma_3 + \sigma_3 \otimes \mathbb{I} ) \mathcal{R}_i$,
with $\mathcal{R}_i = R_i \otimes R_i$, it follows that
\begin{equation}\label{eq14}
    \{\mathbb{I} \otimes \tilde{\sigma}_i + \tilde{\sigma}_i \otimes
    \mathbb{I}\}^{'} = \mathcal{R}_i \{\mathbb{I} \otimes \sigma_3 + \sigma_3 \otimes
    \mathbb{I}\}^{'} \mathcal{R}_i
\end{equation}
and then, after the explicit computation,
\begin{equation}\label{eq15}
    \{ V_i \}^{'} = {\rm span} (\mathbb{I} \otimes \mathbb{I},
    \mathbb{I} \otimes \tilde{\sigma}_i, \tilde{\sigma}_i \otimes
    \mathbb{I}, \tilde{\sigma}_i \otimes \tilde{\sigma}_i, \Omega^+, \Delta_i^-),
\end{equation}
having defined the additional operators
\begin{eqnarray}\label{eq16}
    \Omega^+ &=& \sigma_1 \otimes \sigma_1 + \sigma_2 \otimes \sigma_2 + \sigma_3 \otimes
    \sigma_3, \nonumber \\
    \Delta_i^- &=& \mathcal{R}_i (\sigma_1 \otimes \sigma_2 -  \sigma_2 \otimes
    \sigma_1) \mathcal{R}_i.
\end{eqnarray}
Notice that, in general, the operators in the right hand side of
(\ref{eq15}) are not self-adjoint, nor orthogonal each other in the
Hilbert-Schmidt metric, since the transformation $\mathcal{R}_i$ is
not unitary. However, if the coefficients $\tilde{u}^*_{ij}$, $j =
1, 2, 3$, are real, $\tilde{\sigma}_i$ is self-adjoint and
$\mathcal{R}_i$ unitary (and self-adjoint). Consequently, in this
case the basis of $\{ V_i \}$ is made of Hermitian, orthogonal
operators.

\begin{widetext}
\begin{center}
\begin{table}[t]
  \centering

  \begin{tabular}{|c|c|c|c|c|c|}

    \hline

    Cases & Conditions & $\mathcal{M}$ basis & $\mathcal{M}^{\prime}$ basis & $\mathcal{Z} =
    \mathcal{M} \cap \mathcal{M}^{\prime}$ & $\{ P_n ; n\}$ \\

    \hline

        I &
        $n_+ = 1$, $A = A^T$, $B = A$ &
    \begin{tabular}{c}
        $\mathbb{I} \otimes \mathbb{I}, \mathbb{I} \otimes \tilde{\sigma}_i,
        \tilde{\sigma}_i \otimes \mathbb{I}$, \\
        $\tilde{\sigma}_i \otimes \tilde{\sigma}_i, \Omega^+, \Delta_i^-$
    \end{tabular} &
    \begin{tabular}{c}
        $\mathbb{I} \otimes \mathbb{I}, \tilde{\sigma}_i \otimes \tilde{\sigma}_i$, \\
        $\mathbb{I} \otimes \tilde{\sigma}_i + \tilde{\sigma}_i \otimes \mathbb{I}$
    \end{tabular} &
    $\mathcal{M}^{\prime}$ &
    \begin{tabular}{c}
        $P_1 = \mathcal{R}_i \Pi_1 \mathcal{R}_i$,
        $P_2 = \mathcal{R}_i \Pi_4 \mathcal{R}_i$, \\
        $P_3 = \mathcal{R}_i (\Pi_2 + \Pi_3) \mathcal{R}_i$ \\
    \end{tabular} \\

    \hline

        II &
        $n_+ = 1$, $A \ne A^T$, $B = A$ &
    $\mathbb{I} \otimes \mathbb{I}, \Omega^+$ &
    $\mathcal{M}^{\prime} \supseteq \mathcal{M}$ &
    $\mathcal{M}$ &
    \begin{tabular}{c}
        $P_1 = \Pi_1 + \Pi_4 + \Pi_+$, \\
        $P_2 = \Pi_-$ \\
    \end{tabular} \\

    \hline

        III &
        \begin{tabular}{c}
            \quad $n_+ = n_- = 1$, $A = A^T$, \\
            $B = \alpha A$, $\alpha \in \mathbb{R} \smallsetminus \{ -1, 1 \}$
        \end{tabular} &
    \begin{tabular}{c}
        $\mathbb{I} \otimes \mathbb{I}, \tilde{\sigma}_i \otimes \tilde{\sigma}_i$, \\
        $\mathbb{I} \otimes \tilde{\sigma}_i, \tilde{\sigma}_i \otimes \mathbb{I}$
    \end{tabular} &
    $\mathcal{M}^{\prime} \supseteq \mathcal{M}$ &
    $\mathcal{M}$ &
    \begin{tabular}{cc}
        $P_1 = \mathcal{R}_i \Pi_1 \mathcal{R}_i$, &
        $P_2 = \mathcal{R}_i \Pi_2 \mathcal{R}_i$, \\
        $P_3 = \mathcal{R}_i \Pi_3 \mathcal{R}_i$, &
        $P_4 = \mathcal{R}_i \Pi_4 \mathcal{R}_i$
    \end{tabular} \\

  \hline

      IV &
      $n_+ > 1$, $B = A$ &
  $\mathbb{I} \otimes \mathbb{I}, \Omega^+$ &
  $\mathcal{M}^{\prime} \supseteq \mathcal{M}$ &
  $\mathcal{M}$ &
  \begin{tabular}{c}
      $P_1 = \Pi_1 + \Pi_4 + \Pi_+$, \\
      $P_2 = \Pi_-$ \\
  \end{tabular} \\

    \hline

  \end{tabular}

  \caption{Relevant algebras for the determination of the stationary states
  for the two qubits system, under the assumption $H_S = 0$, in the non-trivial cases.}\label{tab01}
\end{table}
\end{center}
\end{widetext}

The commutants $\{ V_i \}^{'}$ are completely characterized for $i =
1, 2, 3$. Finally, $\{ V_i, V_i^{\dagger} \}^{'}$ can be found by
considering (\ref{eq10}):
\begin{equation}\label{eq15bis}
    \{ V_i, V_i^{\dagger} \}^{'} = \left\{%
\begin{array}{ll}
    \{ V_i \}^{'}, & \hbox{iff $\tilde{\sigma}_i = \tilde{\sigma}_i^{\dagger}$;} \\ \\
    {\rm span} (\mathbb{I} \otimes \mathbb{I}, \Omega^+), \quad & \hbox{otherwise.} \\
\end{array}%
\right.
\end{equation}

The corresponding sets for $i = 4, 5, 6$ can be found by applying
the same procedure to $\hat{\sigma}_i$, assuming that $\lambda^-_i
\ne 0$. The result is
\begin{equation}\label{eq18}
    \{ V_i \}^{'} = {\rm span} (\mathbb{I} \otimes \mathbb{I},
    \mathbb{I} \otimes \hat{\sigma}_i, \hat{\sigma}_i \otimes
    \mathbb{I}, \hat{\sigma}_i \otimes \hat{\sigma}_i, \Omega^-_i, \Delta_i^+),
\end{equation}
where
\begin{eqnarray}\label{eq18bis}
    \Omega^-_i &=& \mathcal{S}_i (\sigma_1 \otimes \sigma_1 - \sigma_2 \otimes
    \sigma_2) \mathcal{S}_i, \nonumber \\
    \Delta_i^+ &=& \mathcal{S}_i (\sigma_1 \otimes \sigma_2 + \sigma_2 \otimes
    \sigma_1) \mathcal{S}_i,
\end{eqnarray}
and $\mathcal{S}_i = S_i \otimes S_i$, with
\begin{equation}\label{eq18bbis}
    \hat{\sigma}_i = \hat{\mu}_i S_i
    \sigma_3 S_i^{-1},
\end{equation}
\begin{equation}\label{eq18bbbis}
    S_i = S_i^{-1} = \frac{1}{\hat{\nu}_i} \left[%
\begin{array}{cc}
  \hat{u}_{i3}^* + \hat{\mu}_i & \hat{u}_{i1}^* - i \, \hat{u}_{i2}^*
  \\ \\
  \hat{u}_{i1}^* + i \, \hat{u}_{i2}^* & - \hat{u}_{i3}^* - \hat{\mu}_i \\
\end{array}%
\right],
\end{equation}
and
\begin{equation}\label{eq13tris}
    \hat{\mu}_i^2 = \sum_j (\hat{u}^*_{ij})^2, \quad \hat{\nu}_i =
    \sqrt{2 \hat{\mu}_i \left( \hat{u}_{i3}^* + \hat{\mu}_i
    \right)}.
\end{equation}
Finally, in this case
\begin{equation}\label{eq18tris}
    \{ V_i, V_i^{\dagger} \}^{'} = \left\{%
\begin{array}{ll}
    \{ V_i \}^{'}, & \hbox{iff $\hat{\sigma}_i = \hat{\sigma}_i^{\dagger}$;} \\ \\
    {\rm span} (\mathbb{I} \otimes \mathbb{I}), \quad & \hbox{otherwise.} \\
\end{array}%
\right.
\end{equation}

If $\tilde{\sigma}_i$ (or $\hat{\sigma}_i$) is singular, the
previous computations are not longer valid. In this case, the
commutants must be evaluated by direct computation and it is not
possible, in general, to express their structure in a compact form.

We have all the ingredients to evaluate the contribution related to
the dissipative generators $V_i$ in (\ref{eq09}) in every situation.
The case in which all the $\lambda_i$ vanish but one has been
discussed above. The remaining cases can be completely described by
considering the following properties.
\begin{enumerate}
\item [(i)] If $\lambda_i^+ \ne 0$ for several indices $i$,
\begin{equation}\label{eq21}
    \bigcap_{i \vert \lambda_i^+ \ne 0} \{ V_i, V_i^{\dagger} \}^{'}
    = {\rm span} (\mathbb{I} \otimes \mathbb{I}, \Omega^+).
\end{equation}
\item [(ii)] If $\lambda_i^- \ne 0$ for several indices $i$,
\begin{equation}\label{eq22}
    \bigcap_{i \vert \lambda_i^- \ne 0} \{ V_i, V_i^{\dagger} \}^{'}
    = {\rm span} (\mathbb{I} \otimes \mathbb{I}).
\end{equation}
\item [(iii)] If $\lambda_i = \lambda_i^+ \ne 0$ and $\lambda_j =
\lambda_{j-3}^- \ne 0$ for a pair of indices $(i, j)$, then
\begin{eqnarray}\label{eq23}
    && \{ V_i, V_i^{\dagger} \}^{'} \cap \{ V_j, V_j^{\dagger} \}^{'}
    = \\ && \left\{%
\begin{array}{ll}
    {\rm span} (\mathbb{I} \otimes \mathbb{I}, \mathbb{I} \otimes \tilde{\sigma}_i,
    \tilde{\sigma}_i \otimes \mathbb{I}, \tilde{\sigma}_i \otimes \tilde{\sigma}_i),
    & \hbox{if $\tilde{\sigma}_i =
    \tilde{\sigma}_i^{\dagger} =\hat{\sigma}_j$;} \\ \\
    {\rm span} (\mathbb{I} \otimes \mathbb{I}), & \hbox{otherwise.} \\
\end{array}%
\right. \nonumber
\end{eqnarray}
\end{enumerate}

To begin with, we assume $H_S = 0$ and we focus on the dissipative
contribution to the dynamics. We denote by $n_+$ and $n_-$ the
number of non-vanishing eigenvalues of the type $\lambda^+$ and
$\lambda^-$ respectively. The relevant algebras in the non-trivial
cases are summarized in Table \ref{tab01}, where the projectors
$\Pi$ are defined as
\begin{eqnarray}\label{eq23bis}
    \Pi_k = [\pi^k_{ij}], \quad \pi^k_{ij} = \delta_{ik}
    \delta_{jk}, \quad k = 1, \ldots 4; \nonumber \\
    \Pi_{-} = \frac{1}{4} (\mathbb{I}\otimes \mathbb{I} - \Omega^+), \quad \Pi_{+}
    = \mathbb{I} \otimes \mathbb{I} - \Pi_{-}.
\end{eqnarray}
We notice that $A = B$ is equivalent to $n_- = 0$. We further
observe that, in case (iii), $[A, B] = 0$, thus it is possible to
choose $\tilde{U} = \hat{U}$. Moreover, $B = \alpha A$ implies
$\tilde{\sigma}_{\xi} = \hat{\sigma}_{\xi}$ for the index ${\xi}$
such that $\lambda_{\xi}^+ \ne 0$ and $\lambda_{\xi}^- \ne 0$.

The sets of projectors defined in Theorem \ref{theo1} are reported
in the table. In the remaining cases $\mathcal{M} = {\rm span}
(\mathbb{I} \otimes \mathbb{I})$, part 1 of Theorem \ref{theo1}
applies and the stationary state is unique. Therefore, the cases
described in Table \ref{tab01} are necessary conditions for the
indirect manipulation of the asymptotic state of the target system
$T$ via the auxiliary system $P$.


\section{Stationary states}
\label{sec2}

We separately explore the non-trivial cases described in Section
\ref{sec1}. Following Theorem \ref{theo1} and considering Table
\ref{tab01}, it is possible to find the family of stationary states
$\rho_S^{\infty}$, and then to extract the corresponding stationary
state of the target subsystem,
\begin{equation}\label{eq24}
    \rho_{T}^{\infty} = {\rm Tr}_{A} \, \rho_S^{\infty},
\end{equation}
by a partial trace over the degrees of freedom of the auxiliary
system. We consider two different choices for the initial state
$\rho_S (0)$. In the spirit of the indirect control scheme, it can
be a factor state,
\begin{equation}\label{eq25}
    \rho_S (0) = \rho_{T} (0) \otimes \rho_{A} (0),
\end{equation}
where $\rho_{T} (0)$ and $\rho_{A} (0)$ are arbitrary states for the
two subsystems, that will be written using a Bloch vector
representation as
\begin{equation}\label{eq26}
    \rho_{T} (0) = \frac{1}{2} \Bigl( \mathbb{I} + \sum_{k = 1}^3 \rho^T_k \sigma_k
    \Bigr),
\end{equation}
and analogously for $\rho_A (0)$, with real coefficients $\rho^T_k$
and $\rho^A_k$. This situation refers to initially uncorrelated
systems, that will in general couple during their joint evolution.
Alternatively, we consider the pure initial state
\begin{equation}\label{eq27}
     \rho_S (0) = \vert \psi \rangle \langle \psi \vert, \quad \vert
     \psi \rangle = \sqrt{P} \vert \uparrow \rangle \otimes \vert
     \uparrow \rangle + \sqrt{1 - P} \vert \downarrow \rangle
     \otimes \vert \downarrow \rangle,
\end{equation}
where $P \in \mathbb{R}$, and $\vert \uparrow \rangle$, $\vert
\downarrow \rangle$ are the $+1$, respectively $-1$ eigenvalues of
the operator $\sigma_3$. This state is entangled if $P \ne 0, 1$,
and it is maximally entangled if $P = \frac{1}{2}$. It is not an
arbitrary entangled state, nevertheless it can show the impact of
initial correlations between the two parties on the manipulation of
the stationary state of $T$.

After discussing the four non-trivial cases presented in Table
\ref{tab01}, we recover the case with $H_S \ne 0$.

\subsection{Case I}
\label{subsec1}

In this case, $\lambda_{\xi} \ne 0$ for some $\xi \in \{1, 2, 3\}$.
For the initial state (\ref{eq25}), the stationary state of the
target system is given by
\begin{eqnarray}\label{eq28}
    \rho^{\infty}_1 &=& u_{\xi 1} \Bigl( \rho^T_1 u_{\xi 1} - \rho^T_2 u_{\xi 2}
    + \rho^T_3 u_{\xi 3} \Bigr) \nonumber \\
    \rho^{\infty}_2 &=& - u_{\xi 2} \Bigl( \rho^T_1 u_{\xi 1} - \rho^T_2 u_{\xi 2}
    + \rho^T_3 u_{\xi 3} \Bigr) \\
    \rho^{\infty}_3 &=& u_{\xi 3} \Bigl( \rho^T_1 u_{\xi 1} - \rho^T_2 u_{\xi 2}
    + \rho^T_3 u_{\xi 3} \Bigr), \nonumber
\end{eqnarray}
where $\rho^{\infty}_k$, $k = 1, 2, 3$, are the Block vector
components of $\rho_T^{\infty}$. They depend solely on the initial
state of $T$, therefore, in this case, it is not possible to
manipulate $\rho_T^{\infty}$ by means of an initially uncorrelated
auxiliary system. However, if the initial state (\ref{eq27}) is
taken into account, a dependence is exhibited in terms of the
Schmidt coefficient $P$:
\begin{eqnarray}\label{eq29}
    \rho^{\infty}_1 &=& (2P - 1) u_{\xi 1} u_{\xi 3} \nonumber \\
    \rho^{\infty}_2 &=& - (2P - 1) u_{\xi 2} u_{\xi 3} \\
    \rho^{\infty}_3 &=& (2P - 1) u_{\xi 3}^2. \nonumber
\end{eqnarray}
Therefore, if it is possible to vary the initial degree of
entanglement, different stationary states for the dynamics result.

\subsection{Case II}
\label{subsec2}

In order to fully characterize the stationary states of the
dynamics, following Theorem \ref{theo1} we need to find a stationary
state $\rho_0$ of maximal rank, such that
\begin{equation}\label{eq30}
    V_{\xi} \rho_0 V_{\xi}^{\dagger} - \frac{1}{2} \{ V_{\xi}^{\dagger} V_{\xi}, \rho_0
    \} = 0,
\end{equation}
where $\xi \in \{1, 2, 3 \}$ satisfies $\lambda_{\xi} \ne 0$.
However, it turns out that every solution of (\ref{eq30}) has at
least a null eigenvalue. In order to prove this, it is not
restrictive to assume that only two coefficients $\tilde{u}^*_{\xi
k}$, $k = 1, 2, 3$, are non-vanishing. In fact, the general case
reduces to this one by a unitary transformation of the Pauli
matrices $\sigma_k$. Since the order is not relevant, we assume that
\begin{equation}\label{eq30bis}
    \tilde{u}^*_{\xi 1} = e^{i \beta_1} \cos{\gamma}, \quad \tilde{u}^*_{\xi 2} = e^{i \beta_2}
    \sin{\gamma}, \quad \tilde{u}^*_{\xi 3} = 0,
\end{equation}
with real $\beta_1$, $\beta_2$ and $\gamma$. The general solution of
(\ref{eq30}) is
\begin{eqnarray}\label{eq30tris}
    \rho_0 &=& \frac{1}{4} \Bigl( \mathbb{I} \otimes \mathbb{I} + r_1 ( \mathbb{I} \otimes
    \sigma_3 + \sigma_3 \otimes \mathbb{I} ) + r_2 \sigma_1 \otimes
    \sigma_1 \\
    &+& r_3 ( \sigma_1 \otimes \sigma_2 + \sigma_2 \otimes \sigma_1
    ) + r_4 \sigma_2 \otimes \sigma_2 + r_5 \sigma_3 \otimes \sigma_3
    \Bigr), \nonumber
\end{eqnarray}
where $a_i$, $i = 1, \ldots, 5$ are real, dependent coefficients,
such that
\begin{eqnarray}\label{eq30tetra}
    r_1 &=& -\frac{1}{2} (4 r_5 + 1) \sin{(\beta_1 - \beta_2) \sin{2 \gamma}}, \nonumber \\
    r_2 &=& \frac{1}{2} \Bigl( 4 r_5 - 1 - (4 r_5 + 1) \cos{2 \gamma}
    \Bigr), \\
    r_3 &=& -\frac{1}{2} (4 r_5 + 1) \cos{(\beta_1 - \beta_2) \sin{2 \gamma}}, \nonumber \\
    r_4 &=& \frac{1}{2} \Bigl( 4 r_5 - 1 + (4 r_5 + 1) \cos{2 \gamma}
    \Bigr). \nonumber
\end{eqnarray}
An explicit computation proves that ${\rm Det} \rho_0 = 0$
irrespective of $r_5$, and then, in full generality, there is not a
maximal rank stationary state, in this case. Therefore, it is not
possible to apply Theorem \ref{theo1}.

\subsection{Case III}
\label{subsec3}

In this case $\rho_0$ is the solution of
\begin{equation}\label{eq31}
    V_{\xi} \rho_0 V_{\xi} + V_{\eta} \rho_0 V_{\eta} -
    \frac{1}{2} \{ V_{\xi}^2 + V_{\eta}^2, \rho_0
    \} = 0,
\end{equation}
where $\xi \in \{1, 2, 3 \}$, $\eta = \xi + 3$ are such that
$\lambda_{\xi}, \lambda_{\eta} \ne 0$, and $V_{\xi}^{\dagger} =
V_{\xi}$, $V_{\eta}^{\dagger} = V_{\eta}$. Since
\begin{eqnarray}\label{eq32}
    V_{\xi} &=& \sqrt{\lambda_{\xi}} \tilde{\mu}_{\xi} \mathcal{R}_{\xi} (\mathbb{I} \otimes \sigma_3 +
    \sigma_3 \otimes \mathbb{I}) \mathcal{R}_{\xi}, \nonumber \\
    V_{\eta} &=& \sqrt{\lambda_{\eta}} \tilde{\mu}_{\xi} \mathcal{R}_{\xi} (\mathbb{I} \otimes
    \sigma_3 - \sigma_3 \otimes \mathbb{I}) \mathcal{R}_{\xi},
\end{eqnarray}
the general stationary state is found to be
\begin{equation}\label{eq33}
    \rho_0 = \frac{1}{4} \Bigl( \mathbb{I} \otimes \mathbb{I} + r_1 \mathbb{I} \otimes \tilde{\sigma}_{\xi}
    + r-2 \tilde{\sigma}_{\xi} \otimes \mathbb{I} + r_3 \tilde{\sigma}_{\xi} \otimes \tilde{\sigma}_{\xi}
    \Bigr),
\end{equation}
where $r_i$, $i = 1, 2, 3$ are real, independent coefficients, and
$r_1 = r_2$ if $\lambda_{\xi} \ne \lambda_{\eta}$. The simplest
choice leading to a maximal rank stationary state is $r_i = 0$ for
all $i$. From it, following Theorem \ref{theo1}, it is possible to
build the complete set of stationary states of the dynamics, and
finally to extract the stationary states of $T$. Although the
algebraic structures are different, it turns out that the results
are the same of Case I, expressed in (\ref{eq28}) and (\ref{eq29}).

\subsection{Case IV}
\label{subsec4}

In this case, the maximal rank stationary state can be found by
solving
\begin{equation}\label{eq34}
    \sum_k \Bigl( V_{k} \rho_0 V_{k}^{\dagger} - \frac{1}{2} \{ V_{k}^{\dagger} V_{k}, \rho_0
    \} \Bigr) = 0,
\end{equation}
where the $V_k$ operators correspond to eigenvalues $\lambda_k^+ \ne
0$, and $k$ assumes two or three values in the set $\{ 1, 2, 3 \}$.
First of all, we notice that $A \ne A^T$ is a necessary condition
for the indirect control of the asymptotic state of $T$. In fact, in
$A = A^T$, all the relevant $V_k$ operators are Hermitian, and then
it is possible to choose the maximal rank stationary state as
$\rho_0 = \frac{1}{4} \mathbb{I} \otimes \mathbb{I}$, leading to
$\rho^{\infty}_k = 0$ for all $k$, for both correlated or
uncorrelated initial states.

For $A \ne A^T$, the general expression of the maximal rank
stationary state $\rho_0$ is rather involved, therefore we prefer to
exhibit a concrete example proving that, in this case, both
uncorrelated and correlated initial states allow indirect
manipulations of the asymptotic states of $T$. We thus assume a
Kossakowski matrix $A$ of the form
\begin{equation}\label{eq35}
    A = \left[%
\begin{array}{ccc}
  a & i d & 0 \\
  -i d & b & 0 \\
  0 & 0 & c \\
\end{array}%
\right],
\end{equation}
where $a, b, c$ and $d$ are real parameters satisfying the
conditions
\begin{equation}\label{eq36}
    \left\{%
\begin{array}{ll}
    a \geqslant 0, \quad b \geqslant 0, \quad c \geqslant 0, \\
    a b - d^2 \geqslant 0, \\
\end{array}%
\right.
\end{equation}
expressing the complete positivity of the evolution. In this case,
the maximal rank stationary state is found to be
\begin{eqnarray}\label{eq37}
    \rho_0 &=& \frac{1}{4} \Bigl( \mathbb{I} \otimes \mathbb{I} + r_1
    (\mathbb{I} \otimes \sigma_3 + \sigma_3 \otimes \mathbb{I}) \\
    &+& r_2 ( \sigma_1 \otimes \sigma_1 - \sigma_2 \otimes \sigma_2 )
    + r_3 \sigma_3 \otimes \sigma_3 \Bigr) \nonumber,
\end{eqnarray}
with
\begin{eqnarray}\label{eq38}
    r_1 &=& \frac{2 d}{a + b} \nonumber \\
    r_2 &=& \frac{(b - a) d^2}{(a + b) (a b + a c + b c)} \\
    r_3 &=& \frac{(a + b + 4 c) d^2}{(a + b) (a b + a c + b c)}. \nonumber
\end{eqnarray}
The asymptotic state of $T$ for the uncorrelated initial state has
components
\begin{eqnarray}\label{eq31}
    \rho^{\infty}_1 &=& 0, \qquad \rho^{\infty}_2 = 0, \\
    \rho^{\infty}_3 &=& \frac{r_1}{3 + 2 r_2 + r_3} \Bigl( 3 + \sum_{k = 1}^3 \rho^T_k \rho^A_k \Bigr); \nonumber
\end{eqnarray}
for the correlated initial state we get
\begin{eqnarray}\label{eq33}
    \rho^{\infty}_1 &=& 0, \qquad \rho^{\infty}_2 = 0, \\
    \rho^{\infty}_3 &=& \frac{4 r_1}{3 + 2 r_2 + r_3} \Bigl( 1 + \sqrt{P (1 - P)} \Bigr). \nonumber
\end{eqnarray}
Therefore, in both cases it is possible to manipulate
$\rho_T^{\infty}$.

\vspace{0.3cm}

Under the assumption $H_S = 0$, the only candidates for the
realization of the asymptotic control protocol by means of an
auxiliary system are the dynamical systems satisfying the conditions
expressed in Case IV, with the additional request $A \ne A^T$. We
now discuss the impact of a non-vanishing $H_S$.

According to (\ref{eq09}), when adding the Hamiltonian term, the
algebras ${\mathcal M}$ computed in Section \ref{sec1} are left
unchanged or reduced, depending on the form of $H_S$. In particular,
in Cases II and IV, they are not modified whenever $[H_S, \Omega^+]
= 0$, that is $H_S$ is invariant under the exchange $T
\leftrightarrow P$ (it contains only terms of the form $\mathbb{I}
\otimes \sigma_i + \sigma_i \otimes \mathbb{I}$ or $\sigma_i \otimes
\sigma_j + \sigma_j \otimes \sigma_i$, with $i, j = 1, 2, 3$).
Otherwise, ${\mathcal M}$ is one-dimensional and part 1 of Theorem
\ref{theo1} applies.

Since in Case III the maximal rank stationary state previously
considered is a stationary state without reference of $H_S$, in
Cases I and III the results obtained under a purely dissipative
dynamics are valid in general. In Cases II and IV, $\rho_0$ can be
evaluated only after specifying $H_S$, therefore it is not possible
to give general results.


\section{Discussion and Conclusions}
\label{sec3}

In this work, we have studied to what extent it is possible to drive
the asymptotic states of a system $T$ via an auxiliary system $P$,
following the indirect control approach, when both target and
auxiliary systems are two-level systems. We have initially
considered the case of a purely dissipative dynamics, and then we
have generalized our results in the presence of an Hamiltonian term.
We have assumed that the two systems interact separately with a
common, homogeneous environment, that is invariant under spatial
translation. This choice is expressed by a particular form of the
Kossakowski matrix $C$ for the composite system. We have found
necessary conditions for the indirect manipulation of the stationary
state of $T$ through the initial state of $P$ when the initial state
is a product state, and provided a concrete example in which this
dependence is indeed apparent. We have also considered the impact of
an initial entanglement between the two systems for control
purposes.

We have found that this kind of control can be performed only when
the blocks of the Kossakowski matrix satisfy the condition $A = B$.
This is not merely a mathematical request; in fact, physical system
that are well described by a quantum dynamical semigroup in this
form can be found in concrete experimental situations, for example
in the study of the resonance fluorescence \cite{arga,puri}, or in
the analysis of the weak coupling of two atoms to an external
quantum field \cite{bena3}, when the distance between the two atoms
can be neglected.

We observe that the phenomenon described in this work has its origin
in the change of the asymptotic behavior of a quantum dynamical
semigroup when the system is enlarged. Consider the example
presented in the previous section, in Case IV. If only the system
$T$ is taken into account, it is described by a dynamical semigroup
with the Kossakowski matrix $A$ given in (\ref{eq35}). Unless $a = b
= d = 0$, this is a uniquely relaxing semigroup with stationary
state
\begin{equation}\label{eq34}
    \rho^{\infty}_1 = 0, \quad \rho^{\infty}_2 = 0, \quad
    \rho^{\infty}_3 = -\frac{2 d}{a + b}.
\end{equation}
However, when adding the probe system, if $A = B$ and $A \ne A^T$,
$S = T + P$ is described by a relaxing semigroup, and the stationary
state is not fixed. In these conditions, multiple stationary states
for $T$ are generated.

The key ingredient for the controllability in indirect control
schemes is given by the correlations between $T$ and $P$. These
correlations can be provided at the beginning, or rather created
during the time evolution. We notice that even a purely dissipative
evolution can provide the needed correlation (for the asymptotic
entanglement in a quantum dynamical semigroup under these
hypotheses, see the results presented in \cite{bena2}). In this
sense, in the indirect control approach the environmental action can
be considered as a resource, not only a source of noise and
decoherence. This kind of behavior has already been observed when
dealing with accessibility and controllability of a pair of qubits
immersed in a bath of decoupled harmonic oscillators, in an exactly
solvable model \cite{roma2}.



\begin{thebibliography}{99}


\bibitem{niel} M.A. Nielsen and I.L. Chuang, {\it Quantum Computation and Quantum
Information}, Cambridge, 2000

\bibitem{shor} P. Shor, SIAM J. Sci. Statist. Comput. 26, 1484 (1997)

\bibitem{benn1} C.H. Bennett, G. Brassard, C. Crepeau, R. Jozsa, A. Peres, and
W.K. Wootters, Phys. Rev. Lett. 70, 1895 (1993)

\bibitem{benn2} C.H. Bennett, F. Bessette, G. Brassard, L. Salvail, and
J. Smolin, Jour. Crypt. 5, 13 (1992)

\bibitem{alic} R. Alicki and K. Lendi, {\it Quantum Dynamical Semigroups and
Applications}, Lecture Notes in Physics, Vol. 286, Springer, Berlin,
1987

\bibitem{breu} H. P. Breuer and F. Petruccione, {\it The Theory of Open Quantum
Systems}, Oxford University Press, Oxford, 2002

\bibitem{lida} D.A. Lidar, K.B. Whaley, {\it Decoherence-Free Subspaces and
Subsystems}, Lecture Notes in Physics, Vol. 622, Springer, Berlin,
2003

\bibitem{tarn} G. M. Huang, T. J. Tarn and J. W. Clark,
J. Math. Phys. 24, 2608 (1983)

\bibitem{rama} V. Ramakrishna, M. V. Salapaka, M. Dahleh, H. Rabitz and
A. Peirce, Phys. Rev. A 51, 960 (1995)

\bibitem{viol1}
L. Viola, S. Lloyd and E. Knill, Phys. Rev. Lett. 83, 4888 (1999)

\bibitem{schi} S. G. Schirmer,
H. Fu and A. I. Solomon, Phys. Rev. A 63, 063410 (2001)

\bibitem{viol2}
S. Lloyd and L. Viola, Phys. Rev. A 65, 010101(R) (2001)

\bibitem{albe}
F. Albertini and D. D'Alessandro, IEEE Transactions on Automatic
Control 48, 1399 (2003)

\bibitem{alta} C. Altafini, J. Math. Phys. 44, 2357 (2003)

\bibitem{bela} V.P. Belavkin, J. Multivariate. Anal. 42, 171 (1992)

\bibitem{wise} H.M. Wiseman and G. J. Milburn, Phys. Rev. Lett. 70, 548
(1993)

\bibitem{manc} S. Mancini, D. Vitali, and P. Tombesi , Phys. Rev. Lett. 80, 688 (1998)

\bibitem{mand} A. Mandilara and J. W. Clark, Phys. Rev. A 71, 013406 (2005)

\bibitem{roma1} R. Romano and D. D'Alessandro, Phys. Rev. A 73, 022323 (2006)

\bibitem{fu} H.C. Fu, H. Dong, X.F. Liu and C.P. Sun, Phys. Rev. A 75, 052317
(2007)

\bibitem{roma3} R. Romano, Phys. Rev. A 75, 024301 (2007)

\bibitem{roma2} R. Romano and D. D'Alessandro, Phys. Rev. Lett. 97, 080402 (2006)

\bibitem{brau} D. Braun, Phys. Rev. Lett. 89, 277901 (2002)


\bibitem{lend} K. Lendi, J. Phys. A 20, 15 (1987)

\bibitem{frig} A. Frigerio, Comm. Math. Phys. 63, 269 (1977)

\bibitem{bena2} F. Benatti and R. Floreanini, Int. J. Quant. Inf. 4, 395 (2006)

\bibitem{arga} G.S. Argawal, A.C. Brown, and G. Vetri, Phys. Rev. A
15, 1613 (1977)

\bibitem{puri} R.R. Puri, {\it Mathematical Methods of Quantum
Optics}, Springer, Berlin, 2001

\bibitem{bena3} F. Benatti and R. Floreanini, J. Opt. B, 429 (2005)



\end{thebibliography}
\end{document}